\documentclass{scrartcl}
\pdfoutput=1
\usepackage[charter]{mathdesign}
\usepackage{helvet}
\usepackage[T1]{fontenc}
\usepackage[utf8]{inputenc}
\usepackage{color}
\usepackage{url}
\usepackage{graphicx}

\makeatletter
\newcommand{\lyxaddress}[1]{
\par {\raggedright #1
\vspace{1.4em}
\noindent\par}
}
\newenvironment{lyxcode}
{\par\begin{list}{}{
\setlength{\rightmargin}{\leftmargin}
\setlength{\listparindent}{0pt}
\raggedright
\setlength{\itemsep}{0pt}
\setlength{\parsep}{0pt}
\normalfont\ttfamily}%
 \item[]}
{\end{list}}

\makeatother

\begin{document}

\title{Model Coupling between the Weather Research and Forecasting Model
and the DPRI Large Eddy Simulator for Urban Flows on GPU-accelerated
Multicore Systems}

\author{Wim Vanderbauwhede}

\date{August-September 2014}

\maketitle

\lyxaddress{School of Computing Science, University of Glasgow, UK\\
In collaboration with Prof. Tetsuya Takemi, DPRI, University of Kyoto,
Japan}
\begin{abstract}
In this report we present a novel approach to model coupling for shared-memory
multicore systems hosting OpenCL-compliant accelerators, which we
call The Glasgow Model Coupling Framework (GMCF). We discuss the implementation
of a prototype of GMCF and its application to coupling the Weather
Research and Forecasting Model and an OpenCL-accelerated version of
the Large Eddy Simulator for Urban Flows (LES) developed at DPRI. 

The first stage of this work concerned the OpenCL port of the LES.
The methodology used for the OpenCL port is a combination of automated
analysis and code generation and rule-based manual parallelization.
For the evaluation, the non-OpenCL LES code was compiled using \emph{gfortran},
\emph{ifort} and \emph{pgfortran}, in each case with auto-parallelization
and auto-vectorization. The OpenCL-accelerated version of the LES
achieves a $7\times$ speed-up on a NVIDIA GeForce GTX 480 GPGPU,
compared to the fastest possible compilation of the original code
running on a 12-core Intel Xeon E5-2640. 

In the second stage of this work, we built the Glasgow Model Coupling
Framework and successfully used it to couple an OpenMP-parallelized
WRF instance with an OpenCL LES instance which runs the LES code on
the GPGPI. The system requires only very minimal changes to the original
code. The report discusses the rationale, aims, approach and implementation
details of this work.
\end{abstract}

\section{Introduction}

\subsection{The Weather Research and Forecasting Model}

The Weather Research and Forecasting Model\footnote{http://www.wrf-model.org}
(WRF) is a mesoscale numerical weather prediction system (NWS) intended
both for forecasting and atmospheric research. It is an Open Source
project, used by a large fraction of weather and climate scientists
worldwide. The WRF code base is written in Fortran-90 and quite complex
and extensive (about 1,000,000 lines of code).

\subsection{The Large Eddy Simulator for the Study of Urban Boundary-layer Flows}

The Large Eddy Simulator for the Study of Urban Boundary-layer Flows
(LES) is developed by Hiromasa Nakayama and Haruyasu Nagai at the
Japan Atomic Energy Agency and Prof. Tetsuya Takemi at the Disaster
Prevention Research Institute of Kyoto University \cite{nakayama2011analysis,nakayama2012large}.
It generates turbulent flows by using mesoscale meteorological simulations,
and was designed to explicitly represent the urban surface geometry.
Its purpose is to conduct building-resolving large-eddy simulations
(LESs) of boundary-layer flows over urban areas under realistic meteorological
conditions. WRF is used to compute the wind profile as input for LES,
so there is a clear case for coupling both models.

\subsection{Model Coupling}

“Model coupling” means using data generated by one model as inputs
for another model: e.g. climate simulations: atmosphere model, ocean
model, land model, ice model. In combination with hardware Acceleration
using GPU/manycore/FPGA, model coupling could result in much reduced
run times and/or higher accuracy simulations. 

Model Coupling is of growing importance because models of e.g climate
need to be as accurate as possible, and therefore include many factors
and effects. Creating a single model incorporating all effects is
increasingly difficult, and requires very large research teams to
cover all specialisms. Combining existing models allows us to model
a large variety of very complex systems 

There are a number of existing libraries to support the process of
model coupling, e.g. MCT, ESMF, OASIS \cite{larson2005model,hill2004architecture,furevik2003description}.
However, each of them requires hand modification of existing model
codes, as well as writing of additional code to control the way the
coupling is done needs to be written as well. Furthermore, current
Model Coupling libraries all use MPI.

\subsection{A Novel Approach to Model Coupling for Multicore/GPU Nodes }

Creating coupled models using the current approach is very difficult
and requires team of experts in various disciplines. As a result model
coupling is too hard for most research teams. A better approach will
allow more researchers to use model coupling and do better science. 

Furthermore, the current approaches were developed for clusters of
single-core machines and are not ideal for multicore/GPU nodes. In
particular, the current approach to load balancing requires that every
model $i$

gets a fixed number of nodes $m_{i}$, proportional to its run time,
i.e. $m_{i}=\alpha\Delta t_{i}$. As a result, data from $m_{i}$
nodes of model $i$ need to be transferred to $m_{j}$ nodes of model
$j$. This is complex and asymmetrical, esp. for more than two models.

\begin{figure}
\begin{centering}
\includegraphics[width=0.8\textwidth]{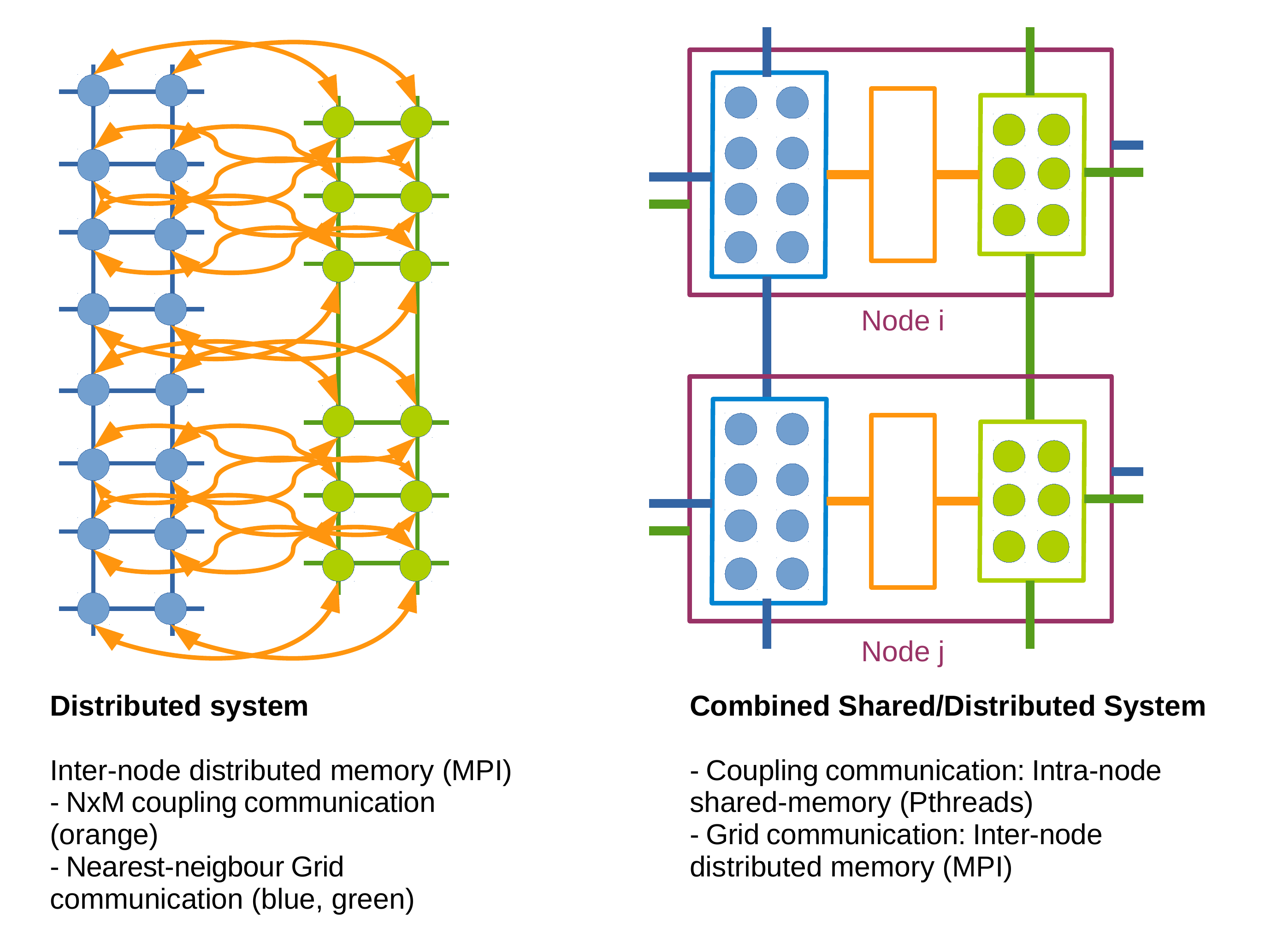}
\par\end{centering}

\captionbelow{\label{fig:Current-and-proposed}Current and proposed approaches
to model coupling}

\end{figure}

Our proposed approach (Fig. \ref{fig:Current-and-proposed}) is to
limit coupling to processes running on a shared-memory node. The nodes
run all processes required for coupling in threads, if a thread is
not needed for coupling it goes to sleep and does not consume CPU
time. As a result, the cluster communication is more symmetrical,
and the load in each node can be balanced more easily.

\section{The OpenCL-Accelerated LES}

The DPRI Large Eddy Simulator (LES) is a high-resolution simulator
which models flow over urban topologies. The main properties of the
DPRI LES are:
\begin{itemize}
\item Generates turbulent flows by using mesoscale meteorological simulations. 
\item Explicitly represents the urban surface geometry.
\item Used to conduct building-resolving large-eddy simulations (LESs) of
boundary-layer flows over urban areas under realistic meteorological
conditions. 
\item Essentially solves the Poisson equation for the pressure, using Successive
Over-Relaxation
\item Written in Fortran-77, single-threaded, about a thousand lines of
code.
\end{itemize}

\subsection{LES Code Structure -- Functional}

The LES structure consists of sequential calls to following subroutines
for each time step:
\begin{description}
\item [{{\small{}velnw:}}] \textsf{\small{}Update velocity for current
time step}{\small \par}
\item [{{\small{}bondv1:}}] \textsf{\small{}Calculate boundary conditions
(initial wind profile, inflow, outflow)}{\small \par}
\item [{{\small{}velfg:}}] \textsf{\small{}Calculate the body force }{\small \par}
\item [{{\small{}feedbf:}}] \textsf{\small{}Calculation of building effects}{\small \par}
\item [{{\small{}les:}}] \textsf{\small{}Calculation of viscosity terms
(Smagorinsky model)}{\small \par}
\item [{{\small{}adam:}}] \textsf{\small{}Adams-Bashforth time integration}{\small \par}
\item [{{\small{}press:}}] \textsf{\small{}Solving of Poisson equation
(SOR)}{\small \par}
\end{description}
\begin{figure}

\begin{centering}
\includegraphics[width=0.6\textwidth]{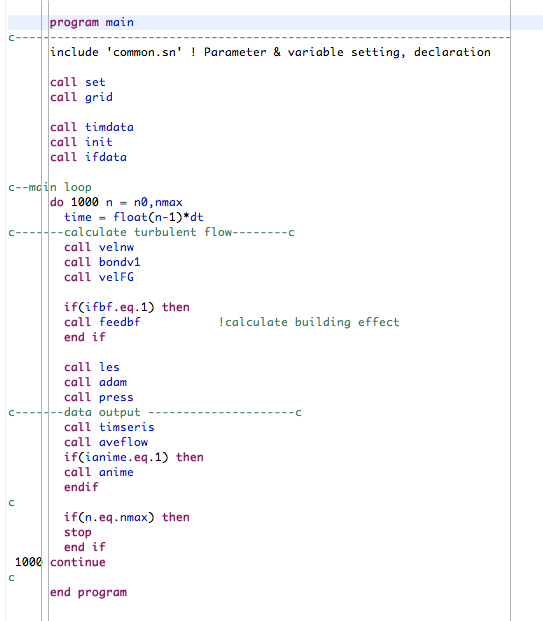}
\par\end{centering}

\captionbelow{LES main program (Fortran-77)}
\end{figure}

\subsection{LES Code Structure -- Computational}

\begin{table}

\begin{centering}
\includegraphics[width=0.7\textwidth]{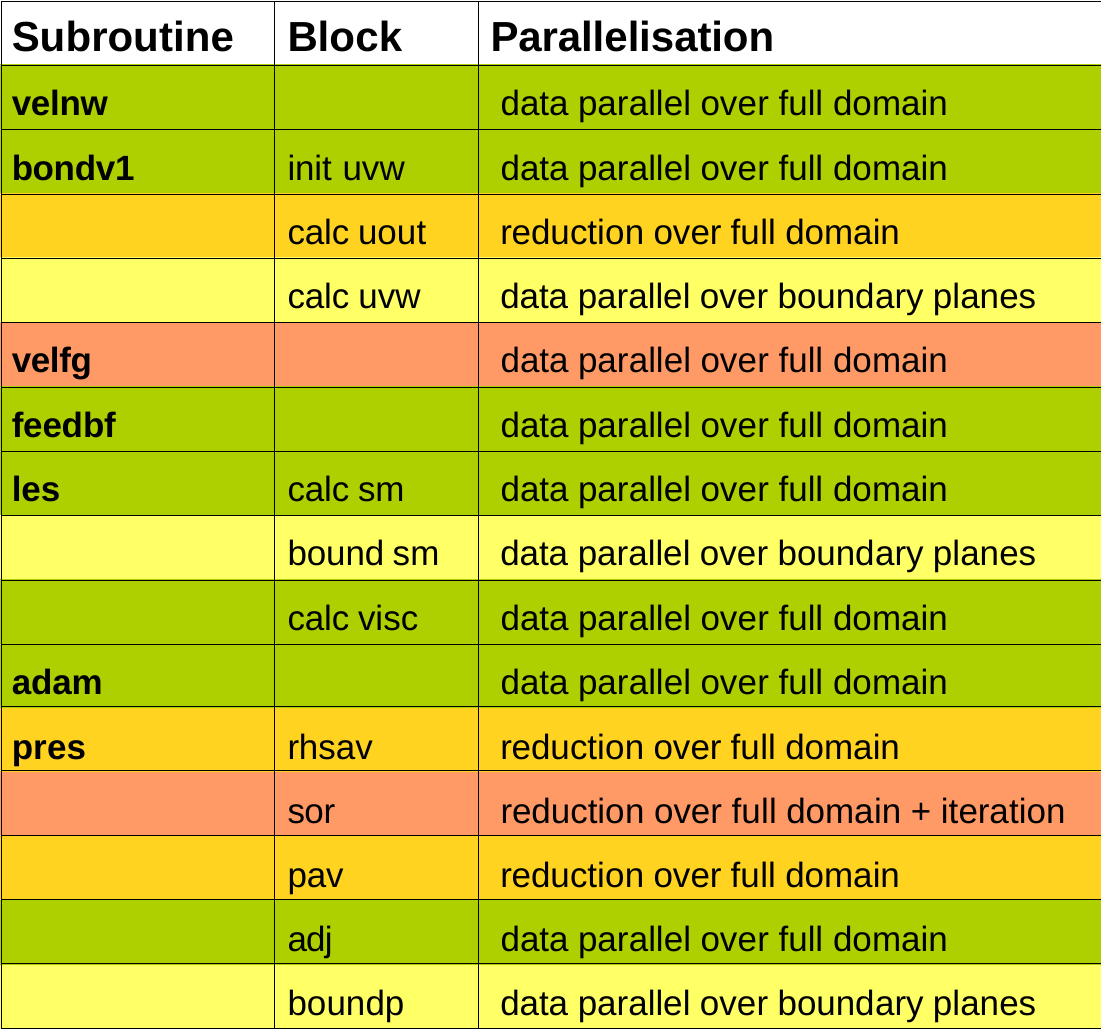}
\par\end{centering}

\captionbelow{LES Computational structure and parallelization strategy for each
block}
\end{table}

\subsection{Methodology}

In this section we discuss our novel methodology for porting Fortran
NWP applications to OpenCL.

\begin{figure}
\begin{centering}
\includegraphics[width=1\textwidth]{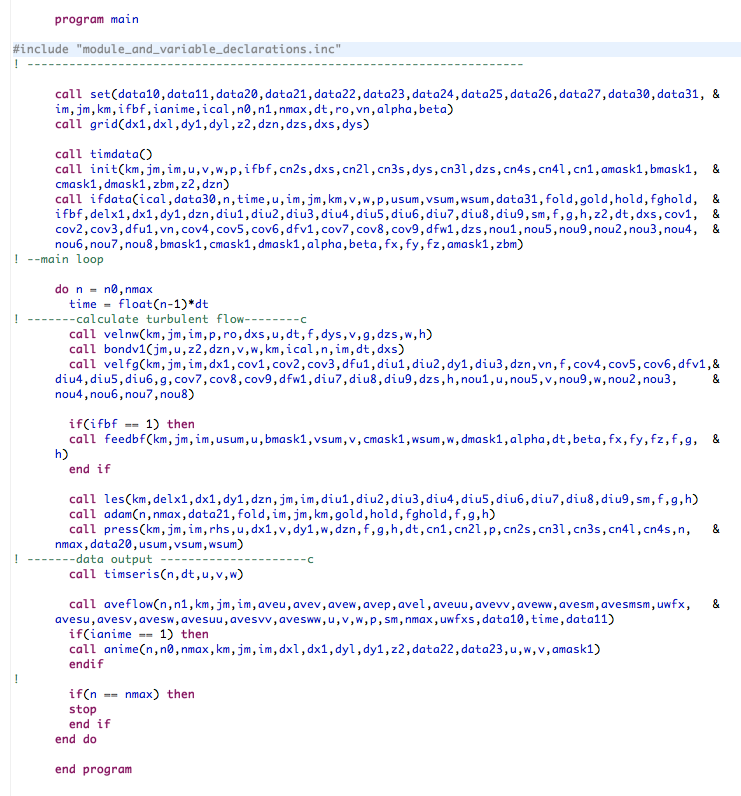}
\par\end{centering}

\captionbelow{LES main program (Fortran-95)}
\end{figure}

The approach used in this work is as follows:
\begin{enumerate}
\item Convert the original F77 code to F95, remove common blocks, if required
refactor into subroutines. All this is done using our \emph{rf4a}
tool. The resulting F95 code has fully explicit subroutine arguments.
\item For each subroutine that could become a kernel, we created a wrapper
in two stages:

\begin{enumerate}
\item We generate a wrapper from the original subroutine using the pragma

\begin{lyxcode}
{\small{}!\$ACC~KernelWrapper(adam)}{\small \par}

{\small{}~~~~~~~call~adam(fgh,fgh\_old,im,jm,km)}{\small \par}

{\small{}!\$ACC~End~KernelWrapper}{\small \par}
\end{lyxcode}

This generates a template for a module \texttt{\small{}module\_adam\_ocl}
which contains a subroutine \texttt{\small{}adam\_ocl}. We manually
edit the generated code if required. 
\begin{itemize}
\item The template file (\texttt{\small{}module\_adam\_ocl\_TEMPL.f95})
is valid Fortran-95, if compiled and run it will be functionally identical
to the original subroutine.
\end{itemize}
\item We generate the OpenCL API from the template using the pragma

\begin{lyxcode}
{\small{}!\$ACC~Kernel(adam\_kernel),~GlobalRange(im{*}jm{*}km),~LocalRange(0)}{\small \par}

{\small{}~~~~~~~~call~adam(fgh,fgh\_old,im,jm,km)~~~~~}{\small \par}

{\small{}!\$ACC~End~Kernel}{\small \par}
\end{lyxcode}
\end{enumerate}

This actually uses even lower-level pragmas underneath, but these
are normally not exposed. The code generator requires a Fortran subroutine
for \texttt{\small{}adam\_kernel}, which initially is a copy original
routine \texttt{\small{}adam}, but can be edited if the OpenCL kernel
arguments would be different.

This generates \texttt{\small{}module\_adam\_ocl.f95} which contains
all necessary calls to OpenCL using our \emph{OclWrapper} library. 
\begin{itemize}
\item If we compile and run the code at this stage it will fail because
the actual OpenCL kernel does not yet exist.
\end{itemize}
\item At this stage we have a wrapper with OpenCL API calls and a ``kernel''
in Fortran, initially a copy of the original subroutine. We convert
this kernel to C using our patched version\footnote{https://github.com/wimvanderbauwhede/RefactorF4Acc/tree/master/F2C-ACC}
of \emph{F2C\_ACC}. Then we further convert this C code to OpenCL,
with some cleaning up. At the moment, this stage is manual, because
the conversion script is not yet ready. 

\begin{itemize}
\item So at this point we have an actual fully working OpenCL version of
the code, albeit a purely sequential one. 
\end{itemize}
\item In this work we manually parallelized the kernels. Ideally, we would
like to use some compiler transformation to assist us with this, esp.
loop merging. 
\item Finally, we merged all kernels into a super-kernel, and also merged
the wrappers. In this work we did this manually but the process is
straightforward to automate.
\end{enumerate}
\begin{figure}

\begin{centering}
\includegraphics[width=1\textwidth]{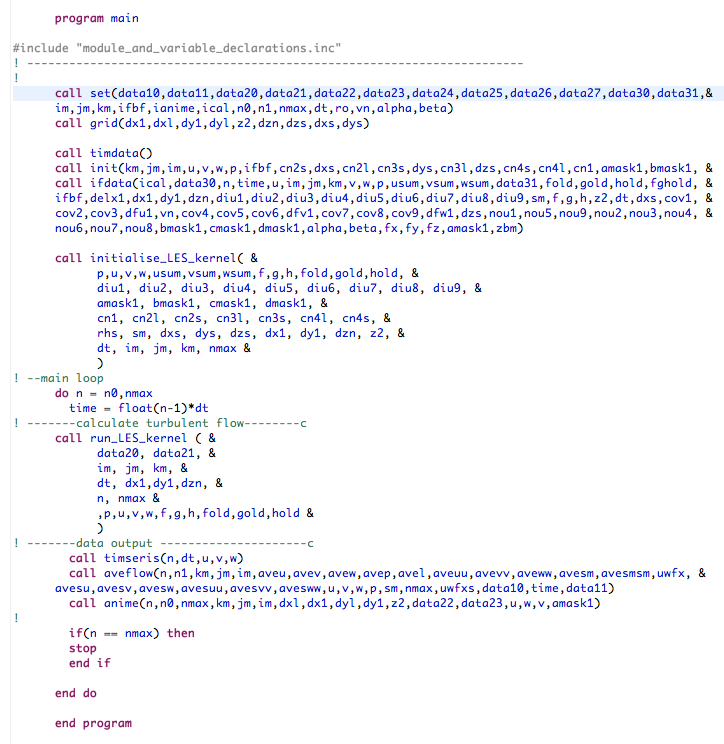}
\par\end{centering}

\captionbelow{LES main program with call to OpenCL kernel wrapper}

\end{figure}

\subsection{Boundary Conditions Computation}

A very effective way to compute boundary conditions in OpenCL is the
\emph{boundary range} approach, where we create a global range consisting
of the sum of the products of the domain dimensions for each direction:
if the domain is $ip\times jp\times kp$ then the boundary range is
$ip\times jp+jp\times kp+ip\times kp$. In this way, the local range
can be set to auto. The kernel has an if-statement to calculate the
different boundary conditions:
\begin{lyxcode}
{\small{}if~(gl\_id~<~jp{*}kp~)~\{}{\small \par}

{\small{}~~unsigned~int~k~=~gl\_id~/~jp;}{\small \par}

{\small{}~~unsigned~int~j~=~gl\_id~\%~jp;}{\small \par}

{\small{}~~//~do~work}{\small \par}

{\small{}\}~else~if~(gl\_id~<~jp{*}kp~+~kp{*}ip)~\{}{\small \par}

{\small{}~~unsigned~int~k~=~(gl\_id~-~jp{*}kp)~/~ip;}{\small \par}

{\small{}~~unsigned~int~i~=~(gl\_id~-~jp{*}kp)~\%~ip;}{\small \par}

{\small{}~~//~do~work}{\small \par}

{\small{}\}~else~if~(gl\_id~<~jp{*}kp~+~kp{*}ip~+~jp{*}ip)~\{}{\small \par}

{\small{}~~unsigned~int~j~=~(gl\_id~-~jp{*}kp~-~kp{*}ip)~/~ip;}{\small \par}

{\small{}~~unsigned~int~i~=~(gl\_id~-~jp{*}kp~-~kp{*}ip)~\%~ip;}{\small \par}

{\small{}~~//~do~work}{\small \par}

{\small{}\}}{\small \par}
\end{lyxcode}
The global range is padded to a multiple of the product of the suggested
number of threads and the number of compute units\footnote{based on CL\_KERNEL\_PREFERRED\_WORK\_GROUP\_SIZE\_MULTIPLE and CL\_DEVICE\_MAX\_COMPUTE\_UNITS},
as this results in much better load balancing:
\begin{lyxcode}
{\small{}unsigned~int~m~=~nthreads{*}nunits;}{\small \par}

{\small{}if~(range~\%~m)~!=~0)~\{}{\small \par}

{\small{}~~range~+=~m~-~(range~\%~m)}{\small \par}

{\small{}\}}{\small \par}
\end{lyxcode}
Consequently, the last branch in the if statement in the kernel must
also be guarded.

\subsection{Some Common Techniques}

\subsubsection{Using OpenCL Vectors for Locality}

The LES uses separate arrays for values in x, y and z directions.
We merged those arrays into float4 arrays, to improve locality and
alignment. Thus, the arrays f,g,h which are actually the values of
the force field f in the x, y and z direction, were merged into fgh.

\subsubsection{Using Private Scalars Instead of Global Arrays}

In several places, the LES computes on global arrays, e.g. the f/g/h
arrays. Where possible, we replaced the global access to fgh(i,j,k)
by a local scalar fgh\_ijk, in effect a form of manual caching.

\subsubsection{Pre-computing Neighboring Points }

The original computation of f/g/h (in the velfg routine) first computes
cov and diu arrays for the full domain, and then accesses them at
neighboring points to compute f/g/h:
\begin{lyxcode}
{\small{}covx1~=~(dx1(i+1){*}cov1(i,j,k)+dx1(i){*}cov1(i+1,j,k))~/(dx1(i)+dx1(i+1))}{\small \par}

{\small{}covy1~=~(cov2(i,j,k)+cov2(i,j+1,k))/2.}{\small \par}

{\small{}covz1~=~(cov3(i,j,k)+cov3(i,j,k+1))/2.}{\small \par}

{\small{}covc~=~covx1+covy1+covz1}{\small \par}

{\small{}dfu1(i,j,k)~=~2.{*}(-diu1(i,j,k)+diu1(i+1,j,k))/(dx1(i)+dx1(i+1))~+~(-diu2(i,j,k)+diu2(i,~\&}{\small \par}

{\small{}j+1,k))/dy1(j)~+~(-diu3(i,j,k)+diu3(i,j,k+1))/dzn(k)}{\small \par}

{\small{}df~=~vn{*}dfu1(i,j,k)}{\small \par}

{\small{}f(i,j,k)~=~(-covc+df)}{\small \par}
\end{lyxcode}
By pre-computing the values for the cov and diu arrays at points i+1/j+1/k+1,
the loops can be merged:
\begin{lyxcode}
{\small{}float~covx1~=~(dx1{[}i+2{]}{*}cov\_ijk.s0+dx1{[}i+1{]}{*}cov\_ijk\_p1.s0)~/(dx1{[}i+1{]}+dx1{[}i+2{]});}{\small \par}

{\small{}float~covy1~=~(cov\_ijk.s1+cov\_ijk\_p1.s1)/2.0F;}{\small \par}

{\small{}float~covz1~=~(cov\_ijk.s2+cov\_ijk\_p1.s2)/2.0F;}{\small \par}

{\small{}float~covc~=~covx1+covy1+covz1;}{\small \par}

{\small{}float~dfu1\_ijk~=~2.0F{*}(-diu\_ijk.s0+diu\_ijk\_p1.s0)/(dx1{[}i+1{]}+dx1{[}i+2{]})~+~(-diu\_ijk.s1+diu\_ijk\_p1.s1)/dy1{[}j{]}~+~(-diu\_ijk.s2+diu\_ijk\_p1.s2)/dzn{[}k+1{]};}{\small \par}

{\small{}float~df~=~vn{*}dfu1\_ijk;}{\small \par}

{\small{}fgh\_ijk.s0~=~(-covc+df);}{\small \par}
\end{lyxcode}

\subsection{SOR Algorithm Implementation}

The major bottleneck of the LES is the solver for the Poisson equation,
which uses Successive Over-Relaxation. The original algoritm is implemented
using the red-black scheme:
\begin{lyxcode}
{\small{}do~nrd~=~0,1}{\small \par}

{\small{}~~do~k~=~1,km}{\small \par}

{\small{}~~~~do~j~=~1,jm}{\small \par}

{\small{}~~~~~~do~i~=~1+mod(k+j+nrd,2),im,2}{\small \par}

{\small{}~~~~~~~reltmp~=~omega{*}(cn1(i,j,k)~{*}(~\&}{\small \par}

{\small{}~~~~~~~~~~cn2l(i){*}p(i+1,j,k)~+~\&}{\small \par}

{\small{}~~~~~~~~~~cn2s(i){*}p(i-1,j,k)~+~\&}{\small \par}

{\small{}~~~~~~~~~~cn3l(j){*}p(i,j+1,k)~+~\&}{\small \par}

{\small{}~~~~~~~~~~cn3s(j){*}p(i,j-1,k)~+~\&}{\small \par}

{\small{}~~~~~~~~~~cn4l(k){*}p(i,j,k+1)~+~\&}{\small \par}

{\small{}~~~~~~~~~~cn4s(k){*}p(i,j,k-1)~-~\&}{\small \par}

{\small{}~~~~~~~~~~rhs(i,j,k))-p(i,j,k))~}{\small \par}

{\small{}~~~~~~~~p(i,j,k)~=~p(i,j,k)~+reltmp}{\small \par}

{\small{}~~~~~~~~sor~=~sor+reltmp{*}reltmp}{\small \par}

{\small{}~~~~~~end~do}{\small \par}

{\small{}~~~~end~do}{\small \par}

{\small{}~~end~do}{\small \par}

{\small{}end~do}{\small \par}
\end{lyxcode}
Conceptually, the p array is divided in red and black points so that
every red point has black nearest neighbors and vice-versa. The new
values for p are computed in two iterations (the \emph{nrd}-loop in
the code example), one for the red, one for the black. 

While this is very effective for single-threaded code, and in fact
also for parallel code on distributed memory systems, it suffers from
poor locality because the accesses to p are strided, and the correction
computation requires access to all six neighbors of p. As a result,
the threads in each compute unit cannot perform coalesced reads or
writes. In\cite{konstantinidis2013graphics}, Konstantinidis and Cotronis
explore a GPU implementation of the SOR method and conclude that their
proposed approach of reordering the matrix elements according to their
color results in considerable performance improvement. However, their
approach is not readily applicable to our problem because one the
one hand we have a 3-D array which is much harder to reorder than
a 2-D array (i.e. the cost of reordering is higher) and also, we cannot
use the reordered array as-is, so we would incur the high reordering
cost twice. 

Therefore, we developed a new technique which we call ``twinned double
buffering'': rather than using the read-black scheme for updating
p, we use a ``double buffer'' update scheme for p: for nrd=0, p\_1
is updated by values from p\_0, and for nrd=1, vice versa. If we would
use two arrays for this, locality would still be poor, so instead
we use an array of vectors of two floats, which we called a ``twinned''
array. Using this data structure and the double buffering scheme,
and assigning the compute units to the k index and the threads in
the compute unit to the i index of the array, we obtain coalesced
memory access. The difference in performance is indeed huge: our approach
is $6\times$ faster than the parallelized version of the red-black
scheme.

\subsection{Evaluation}

In this section we present a validation study of the OpenCL LES by
comparing it to the original Fortran version. We ran the LES on a
domain of 150x150x90 points, with 50 iterations for SOR (default setup).

\subsubsection{Compilers}

The compilers used for the comparison were gfortran 4.8.2 for OpenCL
code, as well as pgf77 12.5-0 and ifort 12.0.0 for the reference code.
We used the following optimizations for auto-vectorization and auto-parallelization:
\begin{itemize}
\item \texttt{\footnotesize{}gfortran -Ofast -floop-parallelize-all -ftree-parallelize-loops=24 }{\footnotesize \par}
\item \texttt{\footnotesize{}pgf77 -O3 -fast -Mvect=simd:256}{\footnotesize \par}
\item \texttt{\footnotesize{}ifort -O3 -parallel}{\footnotesize \par}
\end{itemize}
The run time of the F77 and F95 code is the same with all compilers
(to within a few \%)

\subsubsection{Hardware platforms}

The host platform is an Intel Xeon E5-2640 0 @ 2.50GHz, dual-socket
6-core CPU with two-way hyperthreading (i.e. 24 threads), with AVX
vector instruction support, 32GB memory, 15MB cache, Intel OpenCL
v1.2. The

GPU platform is an NVIDIA GeForce GTX 480 @ 1.40 GHz, 15 compute units,
1.5GB memory, 250KB cache, NVIDIA OpenCL 1.1 (CUDA 6.5.12). 

Table \ref{tab:Hardware-Performance-Indicators} shows the Hardware
Performance Indicators for both systems.

\begin{center}
\begin{table}
\begin{centering}
\includegraphics[width=1\textwidth]{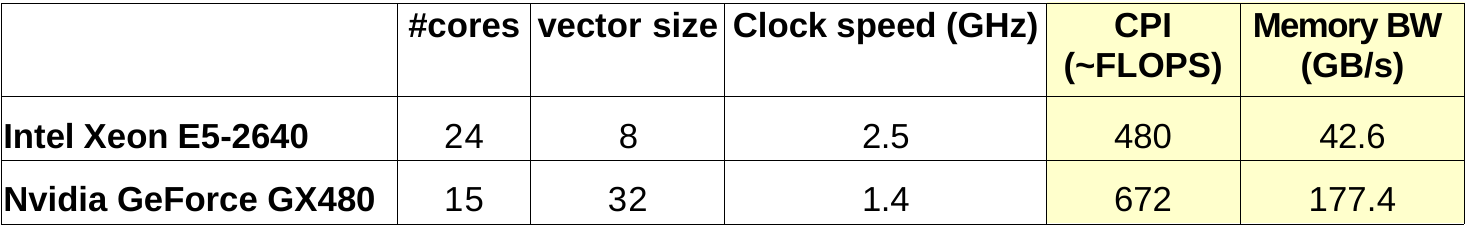}
\par\end{centering}

\captionbelow{Hardware Performance Indicators\label{tab:Hardware-Performance-Indicators}}

\end{table}

\par\end{center}

\subsubsection{OpenCL LES Results}

\begin{figure}
\begin{centering}
\includegraphics[width=0.7\textwidth]{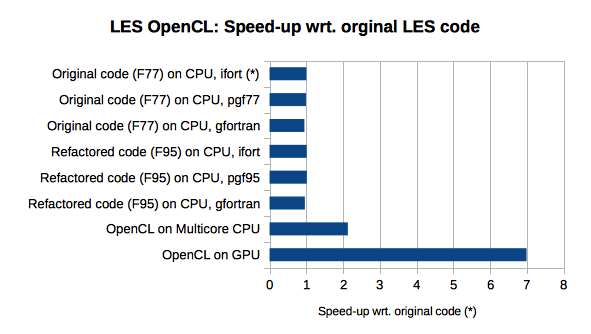}
\par\end{centering}

\captionbelow{Comparison of OpenCL LES with original Fortran code\label{fig:Comparison-of-OpenCL}}
\end{figure}

The results of the comparison of the OpenCL code with the F77 and
F95 reference code results are summarized in Figure \ref{fig:Comparison-of-OpenCL}
. The OpenCL-LES running on the GPU is 7\textbackslash{}times faster
than the fastest reference runtime. 

\begin{figure}

\begin{centering}
\includegraphics[width=0.7\textwidth]{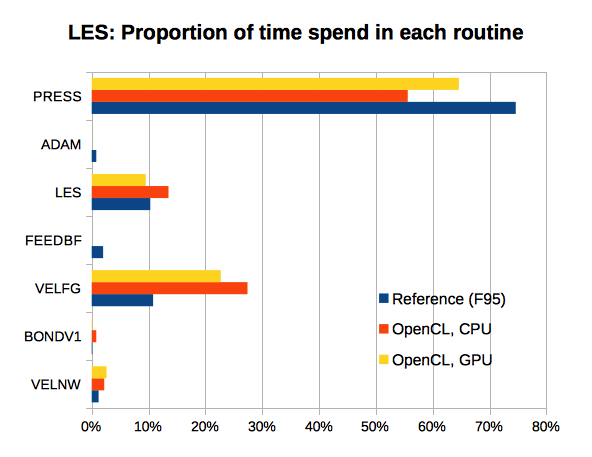}
\par\end{centering}

\captionbelow{Proportion of time spent in each routine\label{fig:Proportion-of-time}}

\end{figure}

\begin{figure}
\begin{centering}
\includegraphics[width=0.8\textwidth]{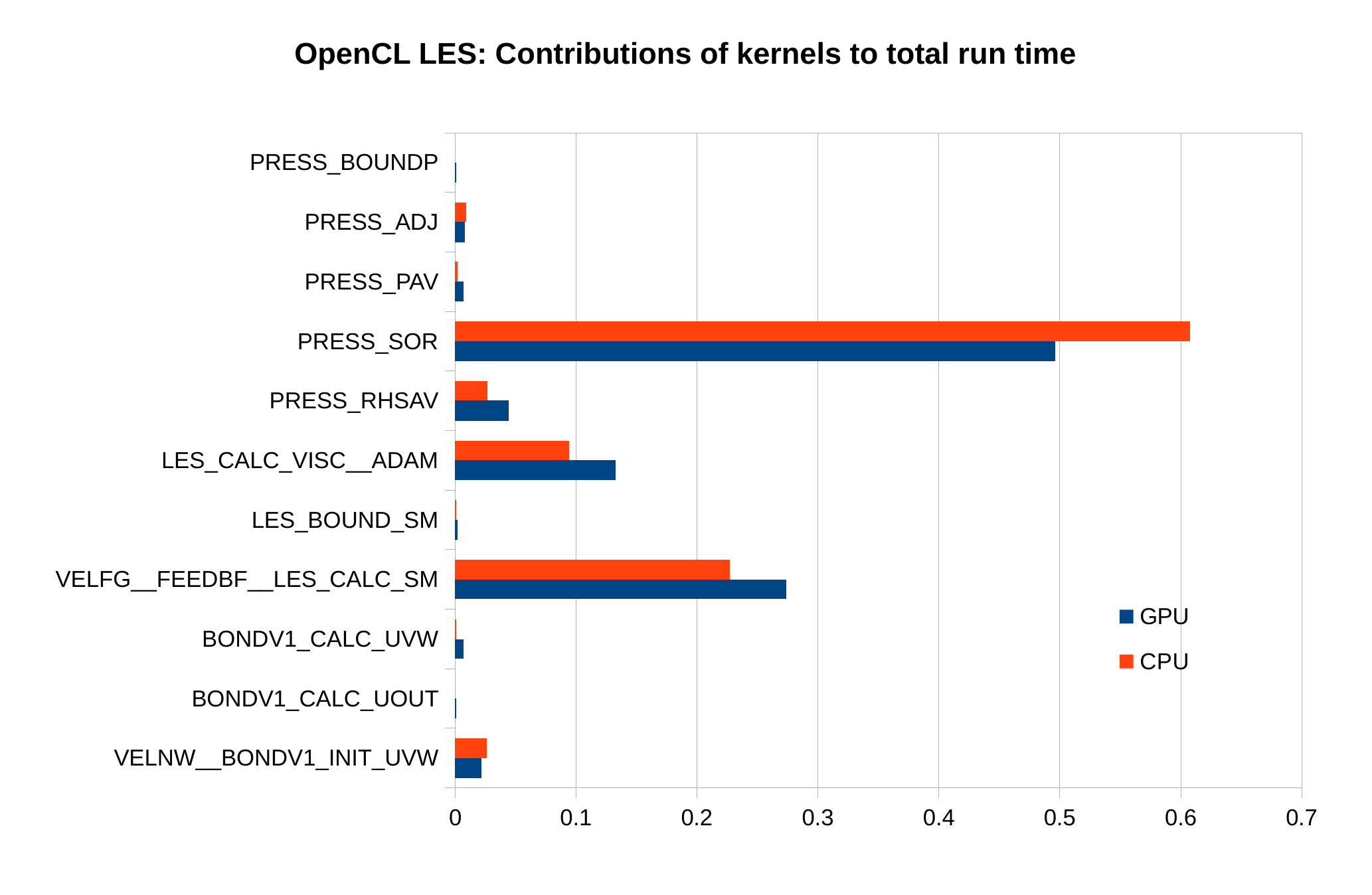}
\par\end{centering}

\label{fig:Proportion-of-time-1}\caption{Proportion of time spent in each OpenCL kernel}
\end{figure}

We implemented all kernels of the LES in OpenCL, because this allows
us to keep all data structures in the GPU memory, rather than copying
between the host and the GPU. Figure \ref{fig:Proportion-of-time}
shows the breakdown of the run time per subroutine in the original
LES code. We see that the press routine, which contains the SOR, takes
most of the run time, followed by the velfg and les routines. Figure
\ref{fig:Proportion-of-time-1} shows the corresponding times for
the actual OpenCL kernels. It should be noted that 50 SOR iterations
is actually on the low side to achieve convergence. So the SOR routine
will dominate the run time entirely for more iterations.

\subsection{OpenCL LES Evaluation Conclusion}

The most important outcome of the effort to convert the LES to OpenCL
is the development of an open-source toolchain to facilitate porting
of Fortran-77 code to OpenCL. Using this toolchain considerable reduces
the porting effort, as illustrated by our work. Furthermore, the porting
of the LES specifically has led to the development of a novel parallel
implementation of the 3-D Successive Over-relaxation algorithm. The
resulting performance is very good: the OpenCL LES running on GPU
is seven times faster than the original code running on a multicore
CPU.

\section{The Glasgow Model Coupling Framework}

In this section we introduce the Glasgow Model Coupling Framework.
We discuss the the longer-term aims of the framework, and status of
the current prototype.

\subsection{Longer-term aims}

Our longer-term aim is to create a system where there is no need for
the end user to write any code at all to achieve model coupling. Instead,
we envisage that our system will use a scenario, written in a formal
way but in a natural language, to express how the models should be
coupled. The scenario will serve as the input to a compilation and
code generation system that will create the fully integrated model
coupling system.

Designing such a specification language is one of the major research
tasks to achieve this goal. We plan to use a system based on Multi-Party
Session Types \cite{ng2012multiparty} to ensure that the scenario
results in a correct system that e.g. is deadlock-free and where data
exchanges between incompatible senders and receivers are impossible.
The Scribble\footnote{http://www.scribble.org/} protocol language
\cite{honda2011scribbling} uses MPST and therefore constitutes a
natural starting point for our investigation. 

However, in order to make the scenarios accessible to a large audience,
we want to explore the use of natural language based specification,
e.g. the Gherkin DSL as used in behavior-driven development \cite{chelimsky2010rspec}.

Furthermore, apart from the scenario, there is also a need for a formal
specification of each model participating in the scenario. One of
the reasons why model coupling is currently very difficult is that
it takes a lot of effort and skill to determine where e.g. the wind
velocity is generated in a given model, what the time loop is, what
the variables and their dimensions are, etc. Our aim is to design
a model specification that allows a code analysis tool to work out
all necessary information -- as opposed to a specification that would
provide all this information, as such as specification would be difficult
and time consuming to produce.

\subsection{GMCF  Model Coupling Operation}

Adopting the classification terminology used by Michalakes \footnote{http://www2.mmm.ucar.edu/wrf/users/workshops/WS2010/presentations/Tutorials/Coupling\%20WRF\%20Michalakes.pdf},
GMCF adopts a component-based coupling model with dataflow-style peer-to-peer
interaction. There is no centralized control or scheduling. Instead,
models communicate using a demand-driven mechanism, i.e. they send
request to other models for data or control information. At each simulation
time step, each model requests time stamps from the models it is coupled
with, and waits until it has received the corresponding time steps.
This synchronization step ensures that the timing of data exchanges
is correct. After syncing, each model can request and/or send data
either before or after computation. When the time loop of a model
finishes, the model broadcast a message to all communicating models.

\subsection{Approach and Status for Current Prototype}

To demonstrate the main concepts of our approach to Model Coupling,
we created a prototype with the express aim to couple WRF and the
DPRI LES model \footnote{https://github.com/wimvanderbauwhede/LES}
using a producer/consumer pattern. As we will see, the same prototype
is readily useable for symmetrical data exchange as well. 

We created a modified version of our Glasgow Parallel Reduction Machine
(GPRM) framework\footnote{https://github.com/wimvanderbauwhede/GannetCode}.
GPRM is a framework for task-based parallel computing, and one can
consider Model Coupling as a special case of this. We call our new
framework the Glasgow Model Coupling Framework (GMCF)\footnote{https://github.com/wimvanderbauwhede/gmcf}.
The necessary modifications are related two different aspects: 
\begin{itemize}
\item GPRM Is a C++ framework, and most scientific models are written in
Fortran. Therefore we developed an automated approach to integrating
Fortran code into GPRM.
\item GPRM uses run-to-completion semantics, i.e. once a task is started,
it runs to completion and only then produces output and checks for
new input data. In a long-running, time-step-base application such
as a NWP model, this approach is not practical, as it would be a huge
effort to lift the the time loop out of the existing model and into
GPRM. Therefore we created a new API which allows communication between
models from inside the model code, obviously essential for model coupling. 
\end{itemize}
Apart from those major changes, we also changed the code organization
and the build system to simplify the creation of a coupled model using
GMCF.

\subsection{More detail on the GMCF architecture}

\subsubsection{Run-time architecture}

\begin{figure}
\begin{centering}
\includegraphics[width=0.8\textwidth]{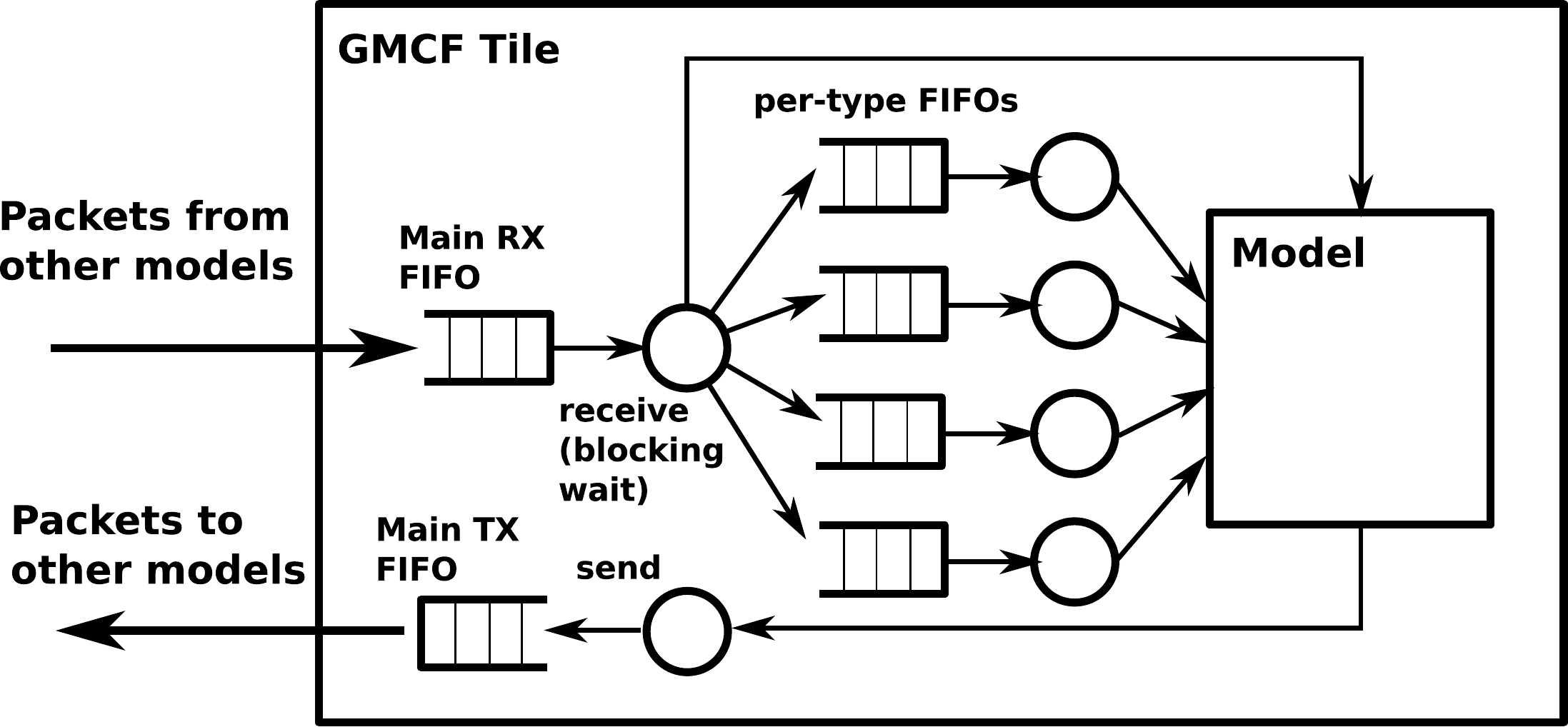}
\par\end{centering}

\label{fig:GMCF-architecture}\caption{GMCF runtime architecture.}

\end{figure}

The run-time architecture of GMCF is illustrated in Fig. \ref{fig:GMCF-architecture}.
On startup, the GMCF runtime system creates a fixed number of threads.
Each of these threads contains a main loop which blocks on a FIFO. 

Every model is allocated to a thread and interfaces with the other
models through a set of FIFOs. The communication between the models
is conceptually based on packets of different types. In other words,
GMCF uses a message-passing approach. As a result, it is in principle
possible to use GMCF for distributed-memory model coupling as well
as shared-memory model coupling. When a packet is read from the main
RX fifo it can be either consumed directly by the model or stored
in a per-type fifo for later use. 

The system is demand-driven: there are request packets and response
packets. The information to exchange between the models is either
time-related, data-related or control-related. 

The low-level GMCF API provides functionality to wait on the main
FIFO, to separate packets out into per-type FIFOs, to shift packets
from the FIFOs and push packets onto the FIFOs. The higher-level API
hides the packet abstraction, instead it expresses synchronization
and data transfer.

\subsubsection{Software architecture}

\begin{figure}
\begin{centering}
\includegraphics[width=0.8\textwidth]{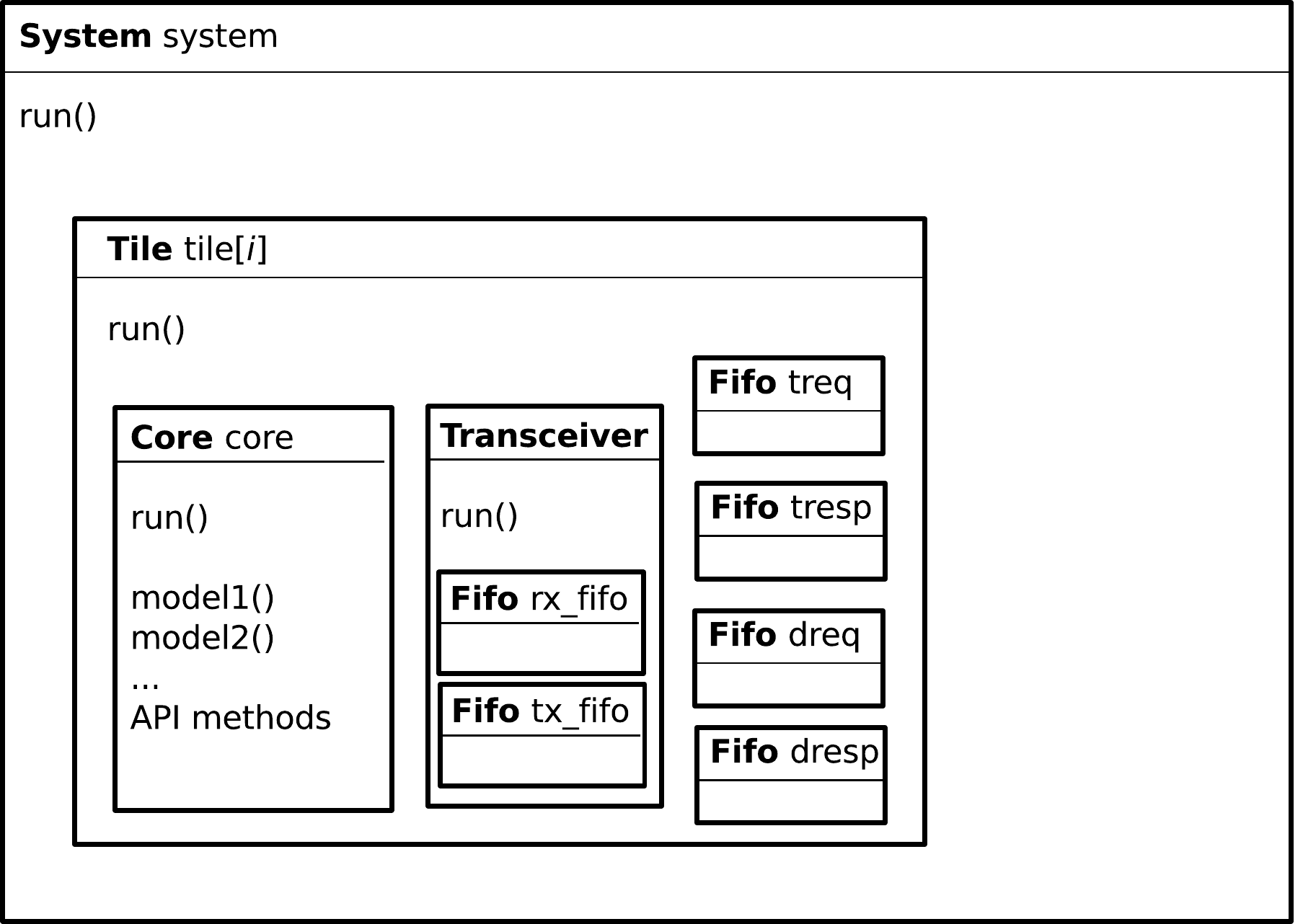}
\par\end{centering}

\label{fig:GMCF-software-architecture}\caption{GMCF software architecture}

\end{figure}

Each of the threads in the GMCF runtime instantiates an object of
the Tile class and calls its run() method, which implements the wait
loop. The Tile class contains the FIFOs and the Core class, which
takes care of the actual model calls (Figure \ref{fig:GMCF-software-architecture}). 

\begin{figure}

\begin{centering}
\includegraphics[width=0.8\textwidth]{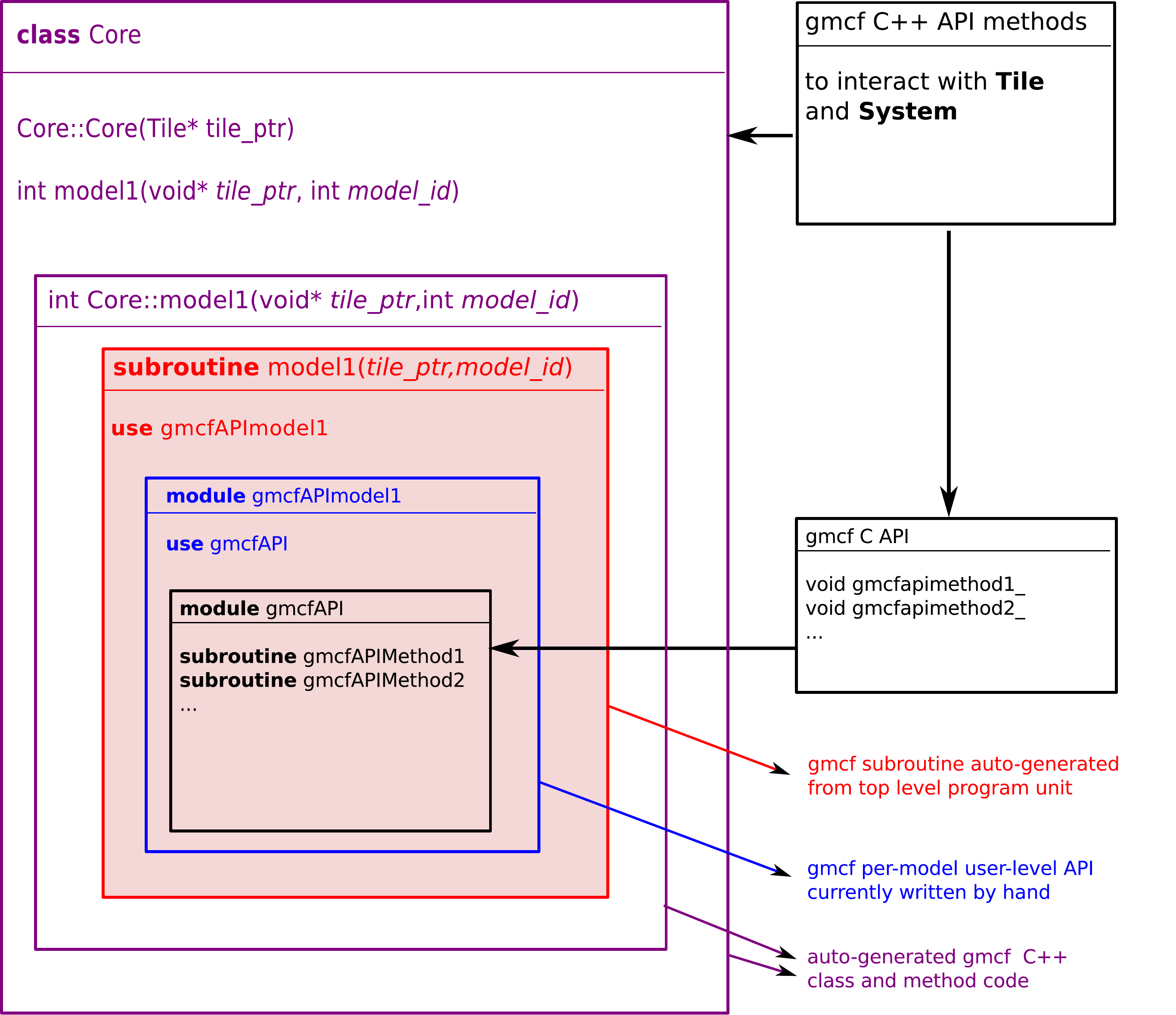}
\par\end{centering}

\captionbelow{\label{fig:GMCF-Fortran-C++-integration}GMCF Fortran-C++ integration
and code generation}

\end{figure}

Our approach is to transform the Fortran top-level program unit into
a subroutine which takes a pointer to the tile as argument (Figure
\ref{fig:GMCF-Fortran-C++-integration}). All model subroutines become
methods of a singleton Core class which inherits from the base Core
class which provides the API calls. This approach allows us to hide
any state in the object. The Core class has a run() method which selects
the model-specific method based on the thread in which it is called
at the start of the run. 

Furthermore, the Fortran API consists of a generic part and a per-model
part, both implemented as modules. The generic part requires the tile
pointer as an argument for each call. The per-model API stores this
pointer in its module so that the final API calls require no additional
arguments.

When the Core.run() method calls the model subroutine, this subroutine
can use the GMCF C++ API method calls to interact with the FIFOs and
so with the other models in the system.

\subsubsection{Code generation architecture}

The main purpose of GMCF is to make model coupling easier. Therefore,
we try to minimize the necessary changes to the original code, and
in fact our long-term aim is to entirely automate the process. Currently,
the user has to manually insert the actual model coupling API calls. 

Apart from that, the build system takes care of the entire integration.
This is less obvious than might seem at first: the GMCF runtime and
API are written in C++, so it is necessary to generate code to interface
with the Fortran model code. The  necessary information required from
the user is very limited: the full path of the top-level program unit
of each model. The build system analyses this program unit and adds
the necessary code for GMCF integration, and generates all required
interface code. 

\section{Coupling WRF and LES}

In this section we present the implementation and preliminary results
of coupling WRF and the OpenCL LES using the GMCF.

WRF is used to compute the wind profile as input for LES. In the original
version of the LES, this input is manually extracted from WRF and
hardcoded in the Fortran source.

\begin{figure}
\begin{centering}
\includegraphics[width=0.7\paperwidth]{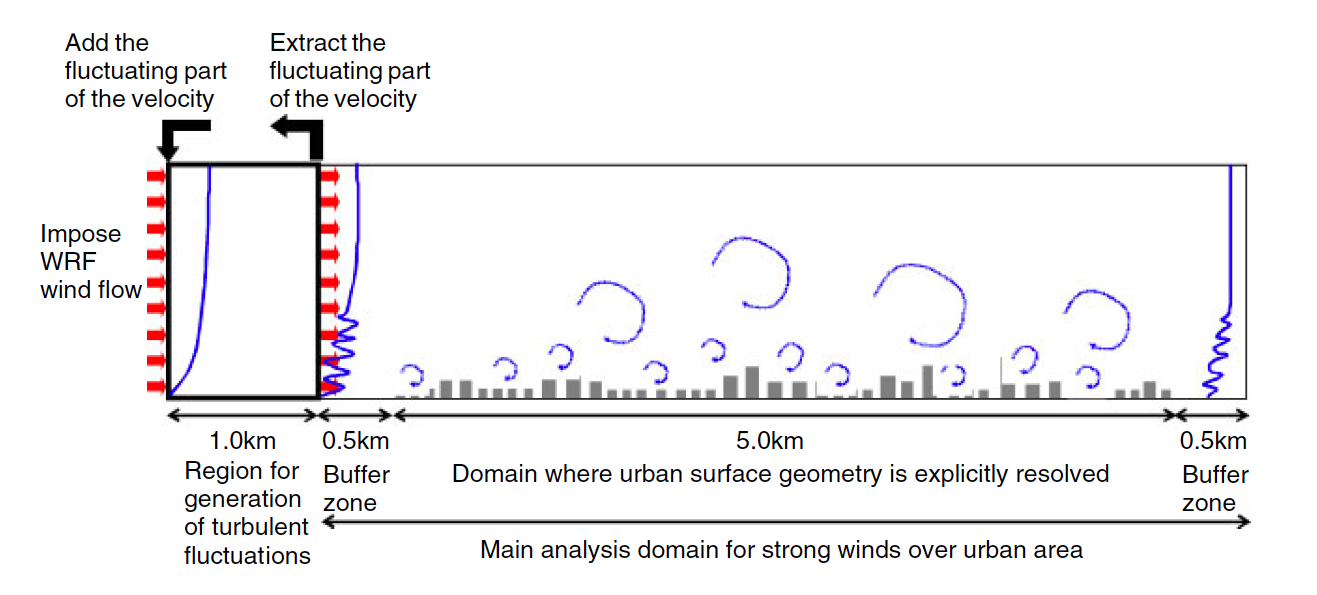}
\par\end{centering}

\captionbelow{\label{Coupling-WRF-and}Coupling WRF and the Large Eddy Simulator}

\end{figure}

\subsection{Implementation Details}

In order to achieve coupling, some modifications are required to both
WRF and LES. On the one hand, the build system needs to be modified
so that the model is compiled into a library rather than an executable.
For LES, as we use SCons, this is trivial. WRF uses a complex build
infrastructure based on make. The changes are still very small, and
some of the changes are actually patches to bugs in the WRF build
system. In fact, the modified WRF build system can now build WRF executables
using both Fortran and C++ main programs. The source code of the main
program unit needs to be modified to make it a subroutine, this is
a very minor change. The other main changes are related to the actual
coupling. Our approach is to create a per-model API based on the generic
GMCF API. Eventually, this per-model API will be automatically generated.
Then the API is used to achieve synchronization and data exchange.

\subsubsection{Creating a gmcf subroutine from the main program unit}

The changes to \texttt{main/wrf.F} and the LES \texttt{main.f95} required
to be able to use GMCF are shown below; these changes are auto-generated
by the build system, the user only needs to provide the name of the
original program and of the new top-level subroutine.
\begin{lyxcode}
\textcolor{blue}{\small{}subroutine~program\_wrf\_gmcf(gmcf\_ptr,~model\_id)}{\small{}~!~This~replaces~'program~wrf'}{\small \par}

\textcolor{blue}{\small{}~~use~gmcfAPI~~~~~~~~~~~~~~~~~~~~~~~~~~~!~gmcf}{\small \par}

\textcolor{blue}{\small{}~~use~gmcfAPIwrf~~~~~~~~~~~~~~~~~~~~~~~~!~gmcf}{\small \par}

{\small{}~~USE~module\_wrf\_top,~only~:~wrf\_init,~wrf\_dfi,~wrf\_run,~wrf\_finalize}{\small \par}

{\small{}~~IMPLICIT~NONE}{\small \par}

{\small{}~}\textcolor{blue}{\small{}~integer(8)~,~intent(In)~::~gmcf\_ptr~~~!~gmcf}{\small \par}

\textcolor{blue}{\small{}~~integer~,~intent(In)~::~model\_id~~~~~~!~gmcf}{\small \par}

\textcolor{blue}{\small{}~~call~gmcfInitWrf(gmcf\_ptr,~model\_id)~~!~gmcf}{\small \par}

{\small{}~~!~Set~up~WRF~model.~~}{\small \par}

{\small{}~~CALL~wrf\_init}{\small \par}

{\small{}~~!~Run~digital~filter~initialization~if~requested.}{\small \par}

{\small{}~~CALL~wrf\_dfi}{\small \par}

{\small{}~~!~WRF~model~time-stepping.~~Calls~integrate().~~}{\small \par}

{\small{}~~CALL~wrf\_run}{\small \par}

{\small{}~~!~WRF~model~clean-up.~~This~calls~MPI\_FINALIZE()~for~DM~parallel~runs.~~}{\small \par}

{\small{}~~CALL~wrf\_finalize}{\small \par}

{\small{}end~}\textcolor{blue}{\small{}subroutine~program\_wrf\_gmcf}{\small \par}

\end{lyxcode}
And for the LES:
\begin{lyxcode}
\textcolor{blue}{\small{}subroutine~program\_les\_gmcf(gmcf\_ptr,~model\_id)}{\small \par}

{\small{}!~...~LES-specific~}\textsl{\small{}use}{\small{}~statements~...}{\small \par}

\textcolor{blue}{\small{}~~~~use~gmcfAPI}{\small \par}

\textcolor{blue}{\small{}~~~~use~gmcfAPIles}{\small \par}

\textcolor{blue}{\small{}~~~~integer(8)~,~intent(In)~::~gmcf\_ptr}{\small \par}

\textcolor{blue}{\small{}~~~~integer~,~intent(In)~::~model\_id}{\small \par}

{\small{}!~...~LES-specific~declarations~...}{\small \par}

\textcolor{blue}{\small{}~~~~call~gmcfInitLes(gmcf\_ptr,~model\_id)}{\small \par}

{\small{}!~...~LES-specific~code~}{\small \par}

\textcolor{blue}{\small{}end~subroutine~program\_les\_gmcf}{\small \par}
\end{lyxcode}

\subsubsection{Time synchronization and data exchange}

The actual synchronization and data exchanges also requires very few
changes, currently these have to be done manually. In WRF, these are
in \texttt{frame/module\_integrate.F}:
\begin{lyxcode}
{\small{}MODULE~module\_integrate}{\small \par}

{\small{}CONTAINS}{\small \par}

{\small{}RECURSIVE~SUBROUTINE~integrate~(~grid~)}{\small \par}

\textcolor{blue}{\small{}~~~~use~gmcfAPI}{\small \par}

\textcolor{blue}{\small{}~~~~use~gmcfAPIwrf}{\small \par}

{\small{}!~...~WRF-specific~code~...}{\small \par}

{\small{}!~WRF~main~time~loop}{\small \par}

{\small{}~~~~~~~~~DO~WHILE~(~.NOT.~domain\_clockisstopsubtime(grid)~)}{\small \par}

\textcolor{blue}{\small{}~~~~~~~~~~~~}\textsf{\textcolor{blue}{\small{}!~Synchronise~simulation~times~between~WRF~\&~LES}}{\small \par}

{\small{}~~~~~~~~~~~}\textcolor{blue}{\small{}~call~gmcfSyncWrf~}{\small \par}

{\small{}!~...~Actual~model~computations~...}{\small \par}

\textcolor{blue}{\small{}~~~~~~~~~~~~}\textsf{\textcolor{blue}{\small{}!~Prepare~the~wind~profile}}{\small \par}

{\small{}~~~~~~~~~~}\textcolor{blue}{\small{}~~call~gmcfCreateWindprofile(grid\%u\_2,grid\%v\_2,grid\%w\_2)}{\small \par}

{\small{}~~~~~~}\textcolor{blue}{\small{}~~~~~~}\textsf{\textcolor{blue}{\small{}!~Send~the~data~to~the~LES~when~requested}}{\small \par}

\textcolor{blue}{\small{}~~~~~~~~~~~~call~gmcfPostWrf}{\small \par}

{\small{}~~~~~~~~~END~DO}{\small \par}

{\small{}~~~~~~}\textcolor{blue}{\small{}~~~call~gmcfFinishedWrf}{\small \par}

{\small{}!~...~WRF-specific~code~...}{\small \par}

{\small{}END~SUBROUTINE~integrate}{\small \par}

{\small{}END~MODULE~module\_integrate~}{\small \par}

\end{lyxcode}
In LES, the time loop is in \texttt{\small{}main.f95}:
\begin{lyxcode}
~~~~{\small{}~~do~n~=~n0,nmax}{\small \par}

{\small{}~~~~~~~~time~=~float(n-1){*}dt}{\small \par}

\textcolor{blue}{\small{}~~~~~~~~}\textsf{\textcolor{blue}{\small{}!~Synchronise~simulation~times~between~WRF~\&~LES}}{\small \par}

\textcolor{blue}{\small{}~~~~~~~~call~gmcfSyncLes}{\small \par}

{\small{}~~~~~~~}\textcolor{blue}{\small{}~}\textsf{\textcolor{blue}{\small{}!~Request~and~receive~the~data~from~WRF}}{\small \par}

\textcolor{blue}{\small{}~~~~~~~~call~gmcfPreLes}{\small \par}

\textcolor{blue}{\small{}~~~~~~~~if~(can\_interpolate~==~0)~then}{\small \par}

\textcolor{blue}{\small{}~~~~~~~~~~~~can\_interpolate~=~1}{\small \par}

\textcolor{blue}{\small{}~~~~~~~~else}{\small \par}

\textcolor{blue}{\small{}~~~~~~~~~~~~}\textsf{\textcolor{blue}{\small{}!~Interpolate~the~value~for~the~current~LES~time~step}}{\small \par}

{\small{}~~~~~~~~}\textcolor{blue}{\small{}~~~~call~gmcfInterpolateWindprofileLes(u,v,w)}{\small \par}

{\small{}!~...~Actual~model~computations}{\small \par}

{\small{}~~~~~~}\textcolor{blue}{\small{}~~end~if~}\textsf{\textcolor{blue}{\small{}!~interpolate~guard}}{\small \par}

{\small{}~~~~~~end~do}{\small \par}

\textcolor{blue}{\small{}~~~~call~gmcfFinishedLes}{\small \par}
\end{lyxcode}
Here the code is a little bit more involved because LES only takes
data from WRF every simulated minute but simulates with a resolution
of half a second. The WRF data are interpolated and the interpolation
can only start after two time steps, hence the guard. 

The LES required other changes as well, because it used a fixed wind
profile rather than taking data from WRF. However, these changes are
not specific to model coupling so they are not included.

\subsubsection{Per-model API}

The changes to the model code are very small because the GMCF operation
is abstracted into a per-model API. This API consists of a small number
of subroutines:
\begin{itemize}
\item The \emph{Init} call initializes GMCF for the given model.
\item The \emph{Sync} call synchronizes the model with the other models
it's communicating with.
\item The \emph{Pre} and \emph{Post} calls are taking care of the actual
data exchange, \emph{Pre} is before computation, \emph{Post} is after
computation. 
\item The \emph{Finished} call informs all other models that the given model
has finished.
\end{itemize}
These subroutines are currently written manually, the next step is
generate the code based on annotations. 
\begin{itemize}
\item For \emph{Init} and \emph{Sync}, what is required is the time step
information relative to some reference, typically the model with the
largest time step. 
\item For the \emph{Pre} and \emph{Pos}t calls, we need to know the names,
types and sizes of the variables containing the data to be exchanged.
\item The \emph{Finished} call can be generated without extra information
\end{itemize}
There are also some API calls that are not as generic as the ones
above: the data received from another model must be assigned to a
variable in the given model, and often we only want to send a portion
of a multidimensional array, so we may want to create intermediate
variables. In the case of WRF and LES, we have \emph{gmcfInterpolateWindprofileLes(u,v,w)}
and \emph{gmcfCreateWindprofileWrf(u,v,w)} but actually the \emph{u,v,w}
are not quite identical in both models. Currently, these subroutines
have to be written by hand. In order to generate the code for these
subroutines, we need to describe exactly how a variable from one model
maps to a variable of another model. Our approach is to create an
intermediate variable for each exchange and define the mapping to
that variable for each model.

\subsection{Building and Evaluation}

Thanks to the extensive code generation and automation, building the
GMCF executable is quite simple: first, specify the models to couple
using a configuration file. This file is used to build the GMCF framework.
Then build each model into a library. Then link the GMCF framework
library and the model libraries into the top-level executable. Detailed
instructions are available on GitHub.

For the evaluation we tested the correctness of the time step synchronization
and the data exchange. And actual evaluation of the performance requires
the WRF input files used for creating the simulation that generates
the wind profile for the LES, and this has not been done yet. 

However, we have successfully built and run a model coupling of WRF
using OpenMP on a multicore GPU with the OpenCL LES running on the
GPU.

\section{Conclusions}

The overall conclusions of this two-month pilot project are very encouraging: 
\begin{itemize}
\item We have demonstrated that our approach to model coupling, aimed at
modern heterogeneous manycore nodes, can be used to couple a complex
NWP simulator such as WRF, parallelized using OpenMP, with a custom
LES simulator running on a GPU using OpenCL. 
\item The LES is a very high-resolution simulator, therefore GPU acceleration
was essential and we have shown that we can achieve a speed-up of
seven times using OpenCL on the GPU. 
\item Furthermore, our model coupling framework is designed to be useable
for automatic generation from scenarios and specifications, and this
will be the focus of our future work. Already, using GMCF requires
only very minor modifications to the original model source code and
build system.
\item All software developed during this project has been open sourced and
is available at \url{https://github.com/wimvanderbauwhede}.
\end{itemize}
This work was conducted during a research visit at the Disaster Prevention
Research Institute of Kyoto University, supported by an EPSRC Overseas
Travel Grant, EP/L026201/1.

\bibliographystyle{plain}
\bibliography{report_model_coupling_OTG2014}

\begin{thebibliography}{1}

\bibitem{chelimsky2010rspec}
David Chelimsky, Dave Astels, Bryan Helmkamp, Dan North, Zach Dennis, and Aslak
  Hellesoy.
\newblock {\em The RSpec book: Behaviour driven development with Rspec,
  Cucumber, and friends}.
\newblock Pragmatic Bookshelf, 2010.

\bibitem{furevik2003description}
Tore Furevik, Mats Bentsen, Helge Drange, IKT Kindem, Nils~Gunnar Kvamst{\o},
  and Asgeir Sorteberg.
\newblock Description and evaluation of the bergen climate model: Arpege
  coupled with micom.
\newblock {\em Climate Dynamics}, 21(1):27--51, 2003.

\bibitem{hill2004architecture}
Chris Hill, Cecelia DeLuca, Max Suarez, Arlindo da~Silva, et~al.
\newblock The architecture of the earth system modeling framework.
\newblock {\em Computing in Science \& Engineering}, 6(1):18--28, 2004.

\bibitem{honda2011scribbling}
Kohei Honda, Aybek Mukhamedov, Gary Brown, Tzu-Chun Chen, and Nobuko Yoshida.
\newblock Scribbling interactions with a formal foundation.
\newblock In {\em Distributed Computing and Internet Technology}, pages 55--75.
  Springer, 2011.

\bibitem{konstantinidis2013graphics}
Elias Konstantinidis and Yiannis Cotronis.
\newblock Graphics processing unit acceleration of the red/black sor method.
\newblock {\em Concurrency and Computation: Practice and Experience},
  25(8):1107--1120, 2013.

\bibitem{larson2005model}
Jay Larson, Robert Jacob, and Everest Ong.
\newblock The model coupling toolkit: A new fortran90 toolkit for building
  multiphysics parallel coupled models.
\newblock {\em International Journal of High Performance Computing
  Applications}, 19(3):277--292, 2005.

\bibitem{nakayama2011analysis}
Hiromasa Nakayama, Tetsuya Takemi, and Haruyasu Nagai.
\newblock Les analysis of the aerodynamic surface properties for turbulent
  flows over building arrays with various geometries.
\newblock {\em Journal of Applied Meteorology and Climatology},
  50(8):1692--1712, 2011.

\bibitem{nakayama2012large}
Hiromasa Nakayama, Tetsuya Takemi, and Haruyasu Nagai.
\newblock Large-eddy simulation of urban boundary-layer flows by generating
  turbulent inflows from mesoscale meteorological simulations.
\newblock {\em Atmospheric Science Letters}, 13(3):180--186, 2012.

\bibitem{ng2012multiparty}
Nicholas Ng, Nobuko Yoshida, and Kohei Honda.
\newblock Multiparty session c: Safe parallel programming with message
  optimisation.
\newblock In {\em Objects, Models, Components, Patterns}, pages 202--218.
  Springer, 2012.

\end{thebibliography}

\end{document}